\documentclass[12pt]{amsart}
\usepackage{amsmath}
\usepackage{amsxtra}
\usepackage{amscd}
\usepackage{amsthm}
\usepackage{amsfonts}
\usepackage{amssymb}
\usepackage{eucal}
\def\({\left(}
\def\){\right)}
\newcommand{\nn}{\nonumber}
\newcommand{\bea}{\begin{eqnarray}}
\newcommand{\ena}{\end{eqnarray}}
\newcommand{\be}{\begin{eqnarray*}}
\newcommand{\en}{\end{eqnarray*}}
\newcommand{\ba}{\begin{array}}
\newcommand{\ea}{\end{array}}

\newcommand{\C}{{\mathbb C}}
\newcommand{\Z}{{\mathbb Z}} 
 
\newcommand{\cP}{\mathcal{P}}

\newcommand{\slt}{\mathfrak{sl}_2}
\newcommand{\res}{{\rm res}}

\newcommand{\tr}{{\rm tr}}

\newcommand{\Tr}{{\rm Tr}}

\renewcommand{\Im}{\mathop{\rm Im}}
\newenvironment{tenumerate}{
  \begin{enumerate}
  
  }{\end{enumerate}}
\newcommand{\bi}{\begin{tenumerate}}
\newcommand{\ei}{\end{tenumerate}}
\newcommand{\isoto}[1][]%
{{\mathop{\buildrel{\sim}\over\longrightarrow}\limits_{#1}}}

\newcommand{\la}{\lambda}

\newcommand{\bs}{\mathbf{s}}

\numberwithin{equation}{section}

\begin{document} 
\title[Homogeneous limit]
{Density matrix of a finite sub-chain of the Heisenberg anti-ferromagnet}
\date{\today}
\author{H.~Boos, M.~Jimbo, T.~Miwa, F.~Smirnov and Y.~Takeyama}
\address{HB: Physics Department, University of Wuppertal, D-42097,
Wuppertal, Germany\footnote{on leave of absence from
Skobeltsyn Institute of Nuclear Physics, MSU, 119992, Moscow, Russia
}}
\email{boos@physik.uni-wuppertal.de}
\address{MJ: Graduate School of Mathematical Sciences, The
University of Tokyo, Tokyo 153-8914, Japan}\email{jimbomic@ms.u-tokyo.ac.jp}
\address{TM: Department of Mathematics, Graduate School of Science,
Kyoto University, Kyoto 606-8502, 
Japan}\email{tetsuji@math.kyoto-u.ac.jp}
\address{FS\footnote{Membre du CNRS}: Laboratoire de Physique Th{\'e}orique et
Hautes Energies, Universit{\'e} Pierre et Marie Curie,
Tour 16 1$^{\rm er}$ {\'e}tage, 4 Place Jussieu
75252 Paris Cedex 05, France}\email{smirnov@lpthe.jussieu.fr}
\address{YT: Graduate School of Pure and Applied Sciences, 
Tsukuba University, Tsukuba, Ibaraki 305-8571, Japan}
\email{takeyama@math.tsukuba.ac.jp}

\begin{abstract}
We consider a finite sub-chain  on an interval
of the infinite XXX model in the ground state. 
The density matrix for such a subsystem
was described in our previous works 
for the model with inhomogeneous spectral parameters. 
In the present paper, we give a compact formula 
for the physically interesting case of the homogeneous model.
\end{abstract}
%%%%%%%%%%%%%%%%%%%%%%%%%%%%%%%%%%%%%
\maketitle
\bigskip

\renewcommand{\Im}{\mathop{\rm Im}}

\setcounter{section}{0}
\setcounter{equation}{0}

\section{Introduction}
The present paper continues 
the study of correlation functions for integrable spin chains
launched in \cite{BJMST1,BJMST2,BJMST3}. 
In our previous works, 
we found an exact expression without involving integrals
for the density matrix of a finite sub-chain of the infinite XXX, XXZ 
and XYZ chains in the ground state. 
More precisely, 
we treated inhomogeneous models in which each site
carries an independent spectral parameter. 
The problem of finding a compact form for the answer in the 
physically important homogeneous case remained open. 
This is exactly the problem which we solve in the present paper.
We shall consider the simplest case of the XXX model.

Consider the isotropic Heisenberg antiferromagnet with the Hamiltonian
$$
H=\frac{1}{2}\sum\limits _{i=-\infty}^{\infty} 
\(\sigma_i^1\sigma_{i+1}^1+ \sigma_i^2\sigma_{i+1} ^2+ \sigma_i^3
\sigma_{i+1}^3\). 
$$
Take a finite sub-chain consisting of sites $i=1,\cdots ,n$. 
The density matrix $\rho_n$  for this sub-chain in the infinite environment 
is an operator acting on $\(\mathbb{C}^2\)^{\otimes n}$. 
Its matrix elements are given by the ground state average
$$
\bigl(\rho _n\bigr)
%{\ }^
^{\epsilon _1,\cdots ,\epsilon _n}_{\bar{\epsilon }_1,\cdots ,\bar{\epsilon} _n}=
\langle \text{vac}|\(E_{\epsilon _1}^{\bar{\epsilon }_1}\)
_1\cdots \(E_{\epsilon _n}^{\bar{\epsilon} _n}\)
_n|\text{vac}\rangle. 
$$
Here $|\text{vac}\rangle$ is the anti-ferromagnetic
ground state, 
$\epsilon_j,\bar{\epsilon}_j=+,-$,   
and 
$\left(E_{\epsilon}^{\bar{\epsilon}}\right)_i$
signifies the matrix unit 
$\left(\delta_{a\epsilon}\delta_{b\bar{\epsilon}}\right)_{a,b=+,-}$ 
acting on the $i$-th tensor component.
It is important to consider a vector $h_n$ belonging to
$\(\mathbb{C}^2\)^{\otimes 2n}$ instead of the matrix $\rho _n$
acting in $\(\mathbb{C}^2\)^{\otimes n}$ . 
%The reason for the necessity of this change 
%from the point of view of physics is unclear to us. 
The vector $h_n$ is given by:
$$
h_n^{\epsilon _1,\cdots ,\epsilon _n, \bar{\epsilon }_n,\cdots ,\bar{\epsilon}_1}
=\prod_{j=1}^n (-\bar{\epsilon}_j)
\cdot \bigl(\rho _n \bigr)^{-\epsilon _1,\cdots ,-\epsilon _n}
_{\ \bar{\epsilon }_1,\ \cdots ,\ \bar{\epsilon} _n}\,.
$$
In the sequel we refer to the tensor components of 
$\(\mathbb{C}^2\)^{\otimes 2n}$ by the indices
$1,\cdots,n,\bar{n},\cdots,\bar{1}$, read from left to right.
The main result of our previous papers can be formulated as follows:
\begin{align}
h_n=
e^{\Omega _n}\bs _n\,.
\label{main}
\end{align}
Here $\bs _n=\prod s_{j,\bar{j}}$ and 
$s_{j,\bar{j}}$ signifies the $\slt$-singlet 
$\frac 1 2\(v_+\otimes v_--v_-\otimes v_+\)$
in the tensor product of two spaces $j,\bar{j}$, 
$v_+,v_-$ being the standard basis of $\C^2$.
The operator
$\Omega _n$ will be defined later (see \eqref{summ}).
It satisfies the condition
\begin{align}
\Omega_n ^{\left[\frac n 2\right]+1}=0
\label{nilp}. 
\end{align}
So, the series for the exponential (\ref{main}) terminates. 
 
\section{Inhomogeneous case}

Let us introduce inhomogeneity parameters $\la _1,\cdots, \la _n$
to the corresponding sites of the lattice 
(see \cite{BJMST1} for more details). 
Then the operator $\Omega_n (\la _1,\cdots ,\la _n)$
becomes dependent on these parameters. 
We define this operator 
following our previous works,  
but we shall slightly change the notation. 

Let $\{S_a\}_{a=1}^3$ be a basis of $\slt$ satisfying 
$[S_a,S_b]=2i\epsilon_{abc}S_c$. 
Following \cite{KRS} define the $L$-operator which belongs to 
$U(\slt)\otimes \text{End}\(\mathbb{C}^2\)$:
\begin{align}
&L(\la)=
\frac{\rho (\la, d)}{\la +\frac d 2 }
L^{(0)}(\la),\quad L^{(0)}(\la)
=
\la +\frac{1}{2}+\frac{1}{2}\sum_{a=1}^3 S_a\otimes\sigma ^a, 
\label{L}
\end{align}
where $d$ is related to  the Casimir operator 
as $\sum_{a=1}^3S_a^2= d^2-1$, and 
$$
\rho (\la,d)=
-
\frac {\Gamma \(\frac 1 2 -\frac  d 4 +\frac {\la } 2\)
\Gamma \(1 -\frac  d 4 -\frac {\la } 2\)}
{\Gamma \(\frac 1 2 -\frac  d 4 -\frac {\la } 2\)
\Gamma \(1 -\frac  d 4 +\frac {\la } 2\)}\,. 
$$
In this normalization 
we have the unitarity and crossing symmetry in the form
$$
L(\la)L(-\la)=1,
\quad 
\sigma^2\left(L(\la)\right)^t\,\sigma^2=-L(-\la-1).
$$
We shall consider
tensor products of several spaces $\mathbb{C}^2$.  
In that case 
the index $i$ in $L_i(\la )$ denotes the 
tensor component as usual. 
In what follows the function $\rho (\la,d)$
always comes in the combination
$$ 
\frac{\rho (\la,d)}{\la+\frac{d}{2}}
\ \frac{\rho (\la -1,d)}{\la+\frac{d}{2}-1}=
-\frac{1}{\la^2-\frac {d^2} 4}\,,
$$
so the $\Gamma$-functions will never really appear. 

We shall also use the ordinary $4\times 4$
$R$-matrix obtained as the image of the 
$L$-operator \eqref{L}
in the 2-dimensional
representation of $U(\slt)$. 
When acting in tensor product of two spaces $i,j$, 
it will be denoted by $R_{i,j}(\la)$. 
We denote the corresponding factor by
$$
\rho (\la):=\rho (\la ,2)\,.
$$

Let us explain the results of the papers \cite{BJMST1,BJMST2} 
in the setting of the XXX model.
First, introduce the operator
\be
T^{[1]}_n(\lambda;\lambda_{2},\ldots,\la_n)
&:=&
L_{\bar2}(\la-\la_2-1)
\cdots L_{\bar n}(\la-\la_n-1)
L_n(\la-\la_n)
\cdots L_2(\la-\la_2)\,.
\en
Notice that in this product the sites $1$, $\bar{1}$ are omitted.

In the paper \cite{BJMST1},  we discussed in detail 
the linear functional on $U(\slt)$ called $\Tr_\la$. 
%This functional is nothing but the analytic continuation of 
%the trace over finite-dimensional irreducible representations 
%with respect to the dimension $d$  of the representation. 
Denote by $\varpi_d$ the irreducible $d$-dimensional 
representation of $U(\slt)$.  
By definition, the map $\Tr_\la:U(\slt)\to\C[\la]$ 
associates with each $A\in U(\slt)$ a unique polynomial 
$\Tr_\la(A)$ of $\la$, such that 
$\Tr_d(A)={\tr }_{\C^d}\varpi_d(A)$   
holds for any positive integer $d$.  
Some of its main properties are 
\be
&&\Tr_\la(AB)=\Tr_\la(BA), \\
&&\Tr_\la\Bigl(\bigl(\sum_{a=1}^3S_a^2\bigr)A\Bigr)
=(\la^2-1)\Tr_\la(A),
\\
&&
\Tr\bigl(e^{tS^3}\bigr)
=\frac{\sinh t \la}{\sinh t}. 
\en
In the present paper we shall also use the 
linear functional
$$
\Tr_{\la_1,\cdots ,\la_k}:\ U(\slt )^{\otimes k}\to \mathbb{C}
[\la_1,\cdots ,\la_k]$$
defined by
$$
\Tr_{\la_1,\cdots ,\la_k}(A_1\otimes\cdots\otimes A_k)=
\prod\limits_{j=1}^k\Tr _{\la_j}(A_j).
$$
%In the paper \cite{BJMST1},  we discussed in detail 
%the linear functional on $U(\slt )$ called $\Tr_d$. 
%This functional is nothing but the analytic continuation of 
%the trace over finite-dimensional irreducible representations 
%with respect to the dimension $d$  of the representation, it is defined
%by the following sequence:
%$$U(\slt )\to U(\slt )/[U(\slt ),U(\slt )]\simeq \mathbb{C}[C]\cdot 1
%\to \mathbb{C}[d^2]\cdot d,$$
%where the last arrow corresponds to evaluation $C=d^2-1$, 
%$\Tr _d (1)=d$.
%In the present paper we shall also use the 
%linear functional
%$$\Tr _{d_1,\cdots ,d_k}:\ U(\slt )^{\otimes k}\to \mathbb{C}
%[d_1,\cdots ,d_k]$$
%defined by
%$$\Tr _{d_1,\cdots ,d_k}(A_1\otimes\cdots\otimes A_k)=
%\prod\limits _{j=1}^k\Tr _{d_j}(A_j).$$

The main ingredient of our construction is the operator
\bea
&&
X_n(\la_1,\cdots,\la_n)_{1,\cdots, n,\bar{n},\cdots ,\bar{1}}
:=
(-1)^{n-1}
\res_{\la =\la _2}
\Tr_{\lambda_{1}-\la}
\Bigr(T^{[1]}_n\bigl(\frac{\lambda_1+\lambda }{2}\bigr)\Bigl)
P_{1,\bar{2}}\cP^-_{1,\bar{1}}\cP^-_{2,\overline{2}}
\,, 
\label{eq:X12}
\ena
where $P_{i,j}$ is the permutation and $\cP ^-_{i,j}=(1-P_{i,j})/2$ is 
the skew-symmetriser. 
Define further
\begin{align}
&
\Omega_n^{(i,j)}(\la_1,\cdots,\la_n)
=-4 \,
\omega (\la _{i,j})X_n^{(i,j)}(\la_1,\cdots,\la_n)\,,
\nn\\
&X_n^{(i,j)}(\la_1,\cdots,\la_n)=
\overleftarrow{\mathbb{R}}_n^{(i,j)}(\la_1,\cdots,\la_n)
\nn\\
&\times
X_n(\la_i,\la_j,\la_1,
\cdots,\widehat{\la_i},\cdots,\widehat{\la_j},\cdots,\la_n)
_{i,j,1\cdots \widehat{i}\cdots \widehat{j}
\cdots n,\bar{n}, \cdots \widehat{\bar j}\cdots\widehat{\bar i}\cdots 
\bar 1,\bar j,\bar i}
\overrightarrow{\mathbb{R}}_n^{(i,j)}(\la_1,\cdots,\la_n), 
\nn
\end{align}
where
\begin{eqnarray*}
&&\overleftarrow{\mathbb{R}}_{n}^{(i,j)}(\la_1,\cdots,\la_n)
\\
&&:=
R_{i,i-1}(\lambda_{i,i-1})\cdots R_{i,1}(\lambda_{i,1}) 
\cdot R_{\overline{i-1},\bar i}(\lambda_{i-1,i})\cdots
R_{\bar 1,\bar i}(\lambda_{1,i}) \\
&&\times 
R_{j,j-1}(\lambda_{j,j-1})\cdots 
R_{j,i+1}(\lambda_{j,i+1})\cdot
R_{j,i-1}(\lambda_{j,i-1})\cdot
\cdots R_{j,1}(\lambda_{j,1})
\\
&&\times R_{\overline{j-1},\bar{j}}(\lambda_{j-1,j})\cdots 
R_{\overline{i+1},\overline{j}}(\lambda_{i+1,j})
\cdot 
R_{\overline{i-1},\overline{j}}(\lambda_{i-1,j})
\cdots R_{\bar{1},\bar{j}}(\lambda_{1,j}),\nn\\
&&\overrightarrow{\mathbb{R}}_n^{(i,j)}(\la_1,\cdots,\la_n)\\
&&:=R _{\overline{n},\overline{i}}(\la _{n,i})\cdots 
R _{\overline{j+1},\overline{i}}(\la _{j+1,i})\cdot
R _{\overline{j-1},\overline{i}}(\la _{j-1,i})
\cdots R_{\overline{i+1},\overline{i}}(\la _{i+1,i})\\
%end of overline
&&\times R _{i,n}(\la _{i,n})
\cdots R _{i,j+1}(\la _{i,j+1})
\cdot R _{i,j-1}(\la _{i,j-1})
\cdots R _{i,i+1}(\la _{i,i+1})\\
&&\times R_{\overline{n},\overline{j}}(\la _{n,j})
\cdots R _{\overline{j+1},\overline{j}}(\la _{j+1,j})
\cdot R_{j,n}(\la _{j,n})\cdots R_{j,j+1} (\la _{j,j+1})
\end{eqnarray*}
%Further define
%$$\Omega ^{(i,j)}(\la_1,\cdots,\la_n)
%=\omega (\la _{i,j})X_n^{(i,j)}(\la_1,\cdots,\la_n),$$
%where
and
$$
\omega (\la )=\frac {d}{d\la }\log \rho (\la )+\frac 1 {2(\la ^2-1)}\,.
$$
The definition of $\Omega ^{(i,j)}$ differs from the one 
used in \cite{BJMST2,BJMST3} \footnote{See eq.(12.1) in 
\cite{BJMST2}. 
We have also used the fact that,  
in the notation there, 
$L_2(\la_{1,2}/2)L_{\bar{2}}(\la_{1,2}/2-1)\cP^-_{2,\bar2}=0$ 
inside the trace and 
$P_{2,\bar2}s_{1,\bar2}s_{\bar{1},{2}}\circ{}_{n-2}\Pi_n
=P_{1,\bar2}\cP^-_{1,\bar1}\cP^-_{2,\bar2}$.}
by the second product of $R$-matrices,  
but the final formula \eqref{main} remains unaltered by this 
modification. 

The operators $\Omega_n^{(i,j)}$ possess
a number of properties the most important among which are
\begin{align}
&\bigl[ \Omega_n^{(i,j)}(\la_1,\cdots,\la_n), \Omega_n^{(k,l)}(\la_1,\cdots,\la_n) \bigr]=0\,,
\label{propO}
\\
&\Omega_n^{(i,j)}(\la_1,\cdots,\la_n)\Omega_n^{(k,l)}(\la_1,\cdots,\la_n) =0
\quad \text{if}\ \{i,j\}\cap\{k,l\}\ne \emptyset\,.
\label{prop1}
\end{align}
As a function of $\la_1,\cdots,\la_n$, 
$\Omega_n^{(i,j)}(\la _1,\cdots ,\la_n)$ is meromorphic.
All poles are simple and located at 
$\la_i,\la _j=\la_l$ and $\la _i,\la _j=\la _l\pm 1$ for $l\ne i,j$
(which are due respectively to $X^{(i,j)}(\la _1,\cdots ,\la_n)$), 
and to the $R$-matrices)
and $\la_{i,j}\in\Z\backslash\{0\}$ (due to $\omega(\la_{i,j})$).
We have
\begin{align}
h_n(\la _1,\cdots ,\la _n)=
e^{\Omega_n(\la _1,\cdots ,\la _n)}\bs _n\,,
\nn
\end{align}
where
\begin{align}
\Omega_n(\la _1,\cdots ,\la_n)=\sum\limits _{1\le i<j\le n}
\Omega_n^{(i,j)}(\la_1,\cdots,\la_n)\,.
\label{summ}
\end{align}
The properties (\ref{propO}), \eqref{prop1} guarantee the nilpotency (\ref{nilp}).

\section{Homogeneous case}

Our goal is to obtain the homogeneous limit 
$\la_1=\cdots=\lambda_n=0$.
In the original formula (\ref{summ}),  
this problem is very complicated:
the singularities 
on the diagonal $\la_i=\la_j$
are present in every term of (\ref{summ}). 
Although these poles are absent in the sum itself, 
it is technically difficult to 
explicitly carry through the cancellation and obtain 
the final answer. 

So, we need to rewrite the formula for 
$\Omega _n(\la _1,\cdots ,\la _n)$
in such a way that taking the homogeneous limit is easier. 
To this end, let us write first of all
another formula for $X_n(\la_1,\cdots,\la_n)$.

Denote by $\varpi _{\la}$ the $\la $-dimensional irreducible representation.
We use only the fact that the Casimir element 
reduces to $\la^2-1$, 
and hence the following computation makes sense 
for non-integer $\la$ as well \cite{BJMST1}.

Notice that 
$$
P^+_a(\la)=
\frac 1 {\la}\varpi _{\la}(L^{(0)}_a(\la /2)),\quad 
P^-_a(\la)=-\frac 1 {\la}\varpi _{\la}(L^{(0)}_a(-\la /2))
$$
are orthogonal projectors.
Consider now $X_n(\la _1,\cdots ,\la _n)$. 
Using the formula $P_{1,\bar 2}\cP ^-_{1,\bar 1}\cP ^-_{2,\bar 2}
=\cP ^-_{1,2}\cP ^-_{\bar 1,\bar 2}P_{1,\bar 2}$,
the definition of 
the $L$-operator and the crossing-symmetry, 
one finds
\begin{align}
\varpi _{\la _{1,2}}\(L^{(0)}_2(\frac{\la _{1,2}}2)
L^{(0)}_1(\frac{\la _{2,1}}2)\)
\cP ^-_{1,2}
&=
{}-\varpi _{\la _{1,2}}\(L^{(0)}_2(\frac{\la _{1,2}}2)
L^{(0)}_2(\frac{\la _{1,2}}2-1)\)
\cP ^-_{1,2}\nn
\\
&=
{}-(\la _{1,2}-1)\varpi _{\la _{1,2}}\(
L^{(0)}_2(\frac{\la _{1,2}}2)\)
\cP ^-_{1,2}\,,
\nn
\\
\varpi _{\la _{1,2}}\(L^{(0)}_{\bar 1}(\frac{\la _{2,1}}2-1)
L^{(0)}_{\bar 2}(\frac{\la _{1,2}}2-1)\)
\cP ^-_{\bar 1,\bar 2}
&=
{}-\varpi _{\la _{1,2}}\(L^{(0)}_{\bar 1}(\frac{\la _{2,1}}2-1)
L^{(0)}_{\bar 1}(\frac{\la _{2,1}}2)\)
\cP ^-_{\bar 1,\bar 2}
\nn
\\ 
&=(\la _{1,2}+1)\varpi _{\la _{1,2}}\(
L^{(0)}_{\bar 1}(\frac{\la _{2,1}}2)\)
\cP ^-_{\bar 1,\bar 2}
\nn\\
&=
{}-(\la _{1,2}+1)\varpi _{\la _{1,2}}\(
L^{(0)}_{\bar 2}(\frac{\la _{1,2}}2-1)\)
\cP ^-_{\bar 1,\bar 2}\,.
\nn
\end{align}

Now it is easy to see that
\begin{align}
&
X_n(\la _1,\cdots ,\la _n)
\label{newX}\\
&
=(-1)^{n-1}
\res_{\mu_1=\la_1}\res _{\mu _2=\la _2}
\ \frac{\mu_{1,2}}
{\mu _{1,2}^2-1}
\Tr_{\mu _{1,2}}
\Bigr(T_n\bigl(\frac{\mu_1+\mu_2}{2}
;\la _1,\cdots, \la _n \bigr)\Bigl)
P_{1,\bar 2}\cP^-_{1,\bar{1}}\cP^-_{2,\overline{2}}\nn
\end{align}
where $T_n(\la)$ is the  complete monodromy matrix:
\be
T_n(\lambda;\la_1,\ldots,\la_n)
&=&
L_{\bar1}(\la-\la_1-1)
\cdots L_{\bar n}(\la-\la_n-1)
L_n(\la-\la_n)
\cdots L_1(\la-\la_1)\,.
\en
Using the 
Yang-Baxter equation one finds the following formula for $X_n^{(i,j)}$:
\begin{align}
X_n^{(i,j)}&(\la_1,\cdots,\la_n)
=(-1)^{n-1}
\res_{\mu_1=\la_i}\res _{\mu _2=\la _j}
\ \frac{\mu_{1,2}}
{\mu _{1,2}^2-1}
\Tr_{\mu _{1,2}}
\Bigr(T_n\bigl(\frac{\mu_1+\mu_2}{2}
;\la _1,\cdots, \la _n \bigr)\Bigl)
\nn
\\
&\times\overleftarrow{\mathbb{R}}_n^{(i,j)}(\la_1,\cdots,\la_n)
P_{i,\bar j}\cP^-_{i,\bar i}\cP ^-_{j,\bar j}
\overrightarrow{\mathbb{R}}_n^{(i,j)}(\la_1,\cdots,\la_n). 
\nn
\end{align}
%Introduce
%\begin{align}
%&t_{n,a}(\la ;\la _1,\cdots ,\la _n):=R_{a,\bar 1}(\la -\la _1-1)
%\cdots R_{a,\bar n}(\la -\la _n -1)R_{a,n}(\la -\la _n)\cdots 
%R_{a,1}(\la -\la _1)\nn
%\end{align}
By a straightforward computation one finds
\begin{align}
&\overleftarrow{\mathbb{R}}_n^{(i,j)}(\la_1,\cdots,\la_n)
P_{i,\bar j}\cP^-_{i,\bar i}\cP ^-_{j,\bar j}
\overrightarrow{\mathbb{R}}_n^{(i,j)}(\la_1,\cdots,\la_n)\nn\\
&=\frac{1}{2}
\Tr _{2,2}\(T_n(\la _i;\la _1,\cdots ,\la _n)\otimes
T_n(\la _j;\la _1,\cdots ,\la _n) \cdot \cP ^-\). \nn
\end{align}
%%%%%%%%%%%%%%%%%%%%%%%%%%%%%%%%%%%%%%%%%%%%%%%
Here the skew-symmetriser $\mathcal{P}^{-}$ acts on the auxiliary space 
$\mathbb{C}^{2} \otimes \mathbb{C}^{2}$. 
Notice that 
$\Tr _{2,2}\(T_n(\mu _1;\la _1,\cdots ,\la _n) \otimes 
T_n(\mu _2;\la _1,\cdots ,\la _n) \cdot \mathcal{P}^{-} \)$ 
is actually symmetric with respect to $\mu _1$, $\mu _2$ due to
the relation
$$
\left[R(\mu),\cP ^-\right]=0.
$$

Obviously, 
the formula for $\Omega _n(\la _1,\cdots ,\la _n)$ can be rewritten now as
\begin{eqnarray} 
{}\qquad 
\Omega _n(\la _1,\cdots ,\la _n)\!\!\!
&=& \!\!\!
\frac{(-1)^n}{2} \!\!\!
\int\!\!\!\int \frac{d\mu_1}{2\pi i}\frac{d\mu_2}{2\pi i}\ 
\omega (\mu_{1,2}) \,
\Tr _{\mu _{1,2}} \!\!
\left( 
T_n \bigl(\frac {\mu _1+\mu _2} 2;\la _1,\cdots ,\la _n\bigr) \right) 
\label{finalf} \\
&& {}\times 
{\rm Tr}_{2,2} 
\Bigl(T_n(\mu _1;\la _1,\cdots ,\la _n)\otimes
T_n(\mu _2 ;\la _1,\cdots ,\la _n) \cdot B(\mu _{1,2})
\Bigr),
\nn
\end{eqnarray}
%%%%%%%%%%%%%%%%%%%%%%%%%%%%%%%%%%%%%%%%%%%%%%%
where 
$$B(\mu _{1,2})=\frac{2 \mu _{1,2}}{\mu _{1,2}^2-1}\cP^{-},$$
the contours of integration encircle the poles $\mu _1 =\la _j$,
$\mu _2 =\la _j$ for $j=1,\cdots ,n$.
The great advantage of this formula is that it allows 
to take the homogeneous limit $\la _j=0$. 
In the next formula we write 
$\Omega _n$ for $\Omega _n(0,\cdots ,0)$, etc.. 
\begin{eqnarray}
\Omega _n&=&
\frac{(-1)^n}{2}
\int\!\!\!\int \frac{d\mu_1}{2\pi i}\frac{d\mu_2}{2\pi i}\ 
\omega (\mu _{1,2})\,
 \Tr _{\mu _{1,2}}\!\!
\left(T_n\bigl(\frac {\mu _1+\mu _2} 2
\bigr) \right) 
\label{finalf2} \\ 
&& \qquad \qquad {}\times 
{\rm Tr}_{2,2}
\Bigl(T_n(\mu _1)\otimes T_n(\mu _2 ) \cdot B(\mu _{1,2})\Bigr),
\nonumber 
\end{eqnarray}
where the integrals are taken around $\mu _i=0$.

Formulas \eqref{finalf}, \eqref{finalf2} are
the main results
of the present paper. Let us
discuss them briefly. 
%First, \eqref{finalf2}
%proves the conjecture of the papers \cite{BK,BKS}. 
%Indeed, 
%since the integrand has a pole of order $n$ at $\mu_j=0$, 
%by evaluating residues  $\Omega _n$ becomes a linear combination of the 
%$\zeta _a(2j-1)$ (the Taylor coefficients of $\omega(\lambda)$) 
%with $j\le n$. 
%Second, formula (\ref{finalf2}) may open up a way for studying 
%the large-distance limit. 
%Also it would be very interesting to see if it helps for 
%the investigation of the limit to continuous field theory. 
%We hope to return to these problems as well as the 
%extension to the XXZ and XYZ cases in future publications.

First we note that the integrand of \eqref{finalf2} 
has a pole of order $n$ at $\mu_j=0$.  
By evaluating residues, 
$\Omega _n$ becomes a linear combination of 
the Taylor coefficients of $\omega(\lambda)$ with $j\le n$,  
given explicitly by 
\begin{eqnarray*}
&&\omega(\lambda)-\frac{1}{2(\lambda^2-1)}
=2\left(\log 2+\sum_{k=1}^\infty\zeta_a(2k+1)\lambda^{2k}\right).
\end{eqnarray*}
Here $\zeta_a(s)=(1-2^{1-s})\zeta(s)$, 
$\zeta(s)$ denoting the Riemann zeta function. 
This settles the conjecture of \cite{BK,BKS}
which states that any correlation function of the XXX model 
can be written as a polynomial of 
$\log 2$ and $\zeta(3),\zeta(5),\cdots$ 
with rational coefficients. 

Second, formula (\ref{finalf2}) may open up a way for studying 
the large-distance limit. 
Also it would be very interesting to see if it helps for 
the investigation of the limit to continuous field theory. 
We hope to return to these problems as well as the 
extension to the XXZ and XYZ cases in future publications.
\bigskip

\noindent
{\it Acknowledgments.}\quad
Research of HB is supported 
by the RFFI grant \#04-01-00352.
Research of MJ is 
supported by 
the Grant-in-Aid for Scientific Research B2--16340033.
Research of TM is
supported by 
the Grant-in-Aid for Scientific Research A1--13304010.
Research of 
FS is supported by INTAS grant \#03-51-3350 and by %\#00-00055 
EC networks  "EUCLID",
contract number HPRN-CT-2002-00325 and "ENIGMA",
contract number MRTN-CT-2004-5652. 
Research of YT is supported by Grant-in-Aid for 
Young Scientists (B) No.\,17740089. 
This work was also supported by the grant of 21st Century 
COE Program at RIMS, Kyoto University. 

HB is grateful to F. G{\"o}hmann, A. Kl{\"u}mper and J. Suzuki 
for discussions.

\end{document}